\documentclass[aps,prl,twocolumn,amsmath,amssymb,floatfix,superscriptaddress]{revtex4-1}
\usepackage{graphicx}
\usepackage{bm}
\usepackage{hyperref}
\usepackage{braket,bbold,color}
\usepackage{tcolorbox}

\newcommand{\bs}[1]{\boldsymbol{#1}}

\begin{document}

\title{Symmetry Indicators for Inversion-Symmetric\\ Non-Hermitian Topological Band Structures}

\author{Pascal M. Vecsei}
\affiliation{Department of Physics, University of Zurich, Winterthurerstrasse 190, 8057 Zurich, Switzerland}

\author{M. Michael Denner}
\affiliation{Department of Physics, University of Zurich, Winterthurerstrasse 190, 8057 Zurich, Switzerland}

\author{Titus Neupert}
\affiliation{Department of Physics, University of Zurich, Winterthurerstrasse 190, 8057 Zurich, Switzerland}

\author{Frank Schindler}
\affiliation{Princeton Center for Theoretical Science, Princeton University, Princeton, NJ 08544, USA}

\date{\today}

\begin{abstract}
We characterize non-Hermitian band structures by symmetry indicator topological invariants. Enabled by crystalline inversion symmetry, these indicators allow us to short-cut the calculation of conventional non-Hermitian topological invariants. In particular, we express the three-dimensional winding number of point-gapped non-Hermitian systems, which is defined as an integral over the whole Brillouin zone, in terms of symmetry eigenvalues at high-symmetry momenta. Furthermore, for time-reversal symmetric non-Hermitian topological insulators, we find that symmetry indicators characterize the associated Chern-Simons form, whose evaluation usually requires a computationally expensive choice of smooth gauge. In each case, we discuss the  non-Hermitian surface states associated with nontrivial symmetry indicators.
\end{abstract}

\maketitle

Non-Hermitian topological band structures generalize the concept of crystalline band structure topology to systems with loss and gain~\cite{NoriNH17,FuNH18,SimonNHSSH18,NoriDirac19,AshvinNH19}. In a first approximation to open quantum systems in condensed matter, the Bloch description of crystals is adapted to accommodate for non-Hermitian hopping matrices, while remaining a Hamiltonian-based single-particle description at heart. The approximation breaks down on time scales comparable to the associated non-Hermitian decay lengths, but provides a first insight into how well-known concepts from Hermitian band theory are modified to account for dissipation. In particular, it turns out that the celebrated bulk-boundary correspondence of topological insulators should be revisited~\cite{Xiong_2018,FloreBiOrthogonal18,YaoZhong18,RJS_NH20, stegmaier2020topological, yang2020non, yang2021fermion}: the non-Hermitian skin effect~\cite{HatanoNelson96,HatanoNelson97,HatanoNelson98,Lee:2016,Torres:2018,LonghiSkin19,YaoZhongSkin19,Lee:2019,Lee:2019,TopoSkin20} is a striking example of the sensitive dependence of (single-particle~\cite{BohmSkinPauli20,CHL_RSFermi_20}) non-Hermitian systems on the boundary conditions. In tradition with many other topological features of electronic band structures, it has already been theoretically predicted and experimentally realized in a range of classical analogue systems~\cite{Helbig20Skin,HofmannReciprocal20,HatsugaiSkin20,LiCriticalSkin20,Ghatak29561, weidemann2020topological}.

Nevertheless, the bulk classification of topological insulators with respect to (crystalline) symmetries can be adapted rather straightforwardly to non-Hermitian systems, as long as the notion of bulk gap is specified~\cite{UedaPaper18,KawabataClassification19,NHReflectionClassChen19,NHDefectClass19,HarryPeriodicTable19}: for Hermitian insulators, the gap separates the occupied and empty band subspaces. For non-Hermitian systems, which in general have complex energy spectra, one distinguishes line gaps and point gaps. In the complex plane, a line gap separates the spectrum into two disconnected ``bands", while a point gap constitutes a region (centered around $E=0$ without loss of generality) that is devoid of, but fully surrounded by, eigenstates. Only point-gapped systems are intrinsically non-Hermitian, in that they cannot be deformed to Hermitian systems without closing the gap.

A (partial) topological classification of point-gapped insulators has been achieved, but the corresponding topological invariants are often formulated in terms of computationally expensive Brillouin zone integrals~\cite{UedaPaper18,KawabataClassification19, zhang2020correspondence}. For Hermitian systems, crystalline symmetries have often been fruitfully used to simplify the calculation of topological invariants of non-crystalline topological phases~\cite{FuKane07,Slager17,Po_2017,Khalaf17,Bradlyn17,Ono18}. In this Letter, we show that the same is possible for non-Hermitian band structures: We first prove symmetry indicator formulas for the winding numbers of one-dimensional (1D) and three-dimensional (3D) non-Hermitian insulators in Altland-Zirnbauer class A. Then, we formulate a symmetry-indicator based invariant for 3D point-gapped systems in symmetry class AII, where the usual invariant is especially costly in that it requires the choice of a smooth gauge over the Brillouin zone~\cite{ChiuReview16}. Our results point towards a unified understanding of Hermitian and non-Hermitian crystalline topology.

\emph{1D winding number---}Consider the disorder-free Hatano-Nelson chain~\cite{HatanoNelson96,HatanoNelson97,HatanoNelson98} with Hamiltonian
\begin{equation}
H = \sum_i t_\mathrm{r} c^\dagger_{i+1}c_i + t_\mathrm{l} c^\dagger_i c_{i+1},
\end{equation}
where $t_\mathrm{r} \neq t_\mathrm{l}$ are the real-valued right and left hopping amplitudes, respectively, and $c^\dagger_i$ creates an electron at site $i \in 1 \dots L$ ($L$ is the system size, we choose a lattice spacing $a=1$). With periodic boundary conditions (PBC), the system is described by the $1 \times 1$ Bloch Hamiltonian
\begin{equation} \label{eq: HatanoNelsonBloch}
\mathcal{H}(k) = t_\mathrm{r} e^{\mathrm{i} k} + t_\mathrm{l} e^{-\mathrm{i} k},
\end{equation}
where $k = 2\pi/L \dots 2\pi$ is the crystal momentum. The spectrum then forms an ellipse in the complex plane, and the single-particle eigenstates are standard Bloch waves. Under the introduction of open boundary conditions (OBC), the spectrum collapses onto the real line, while all eigenstates accumulate on only one edge of the system. For $t_\mathrm{r} > t_\mathrm{l}$, this is the right edge, while for $t_\mathrm{r} < t_\mathrm{l}$, this is the left edge. This distinctly non-Hermitian phenomenon of spectral collapse in OBC is dubbed the non-Hermitian skin effect~\cite{HatanoNelson96,HatanoNelson97,HatanoNelson98,Lee:2016,Torres:2018,Xiong_2018,YaoZhong18,FloreBiOrthogonal18,LonghiSkin19,YaoZhongSkin19,Lee:2019,TopoSkin20,RJS_NH20}.

Importantly, the skin effect is an unavoidable property of all Bloch Hamiltonians that have a nontrivial 1D winding number (taking $L \rightarrow \infty$)
\begin{equation}
w_{\mathrm{1D}} = \int_{0}^{2\pi} \frac{\mathrm{d}k}{2\pi \mathrm{i}} \frac{\mathrm{d}}{\mathrm{d} k} \mathrm{log}\,\mathrm{det}\,\mathcal{H}(k) \in \mathbb{Z},
\end{equation}
which is well-defined in presence of a point gap at $E = 0$~\cite{TopoSkin20}. For the Hamiltonian in Eq.~\eqref{eq: HatanoNelsonBloch}, we obtain $w_{\mathrm{1D}} = \mathrm{sgn}(t_\mathrm{r}-t_\mathrm{l})$. In this sense, $w_{\mathrm{1D}}$ is an integer-valued invariant of intrinsically 1D non-Hermitian topological phases that requires no symmetries for its quantization. (The topological classification of non-Hermitian systems in Altland-Zirnbauer class A is $\mathbb{Z}$~\cite{KawabataClassification19}). 

In the presence of crystalline inversion symmetry, $w_{\mathrm{1D}} \mod 2$ can be expressed purely in terms of inversion eigenvalues at inversion-symmetric momenta. For this, first note that for non-Hermitian systems, there are two ways of implementing inversion symmetry: (1) ``standard" inversion symmetry implies
\begin{equation}
\mathcal{I} \mathcal{H}(k)\mathcal{I}^\dagger = \mathcal{H}(-k),
\end{equation}
for some unitary matrix $\mathcal{I}$. (2) ``Pseudo" inversion symmetry implies 
\begin{equation} \label{eq: PseudoInversionConstraint}
\mathcal{I} \mathcal{H}(k)\mathcal{I}^\dagger = \mathcal{H}^\dagger(-k).
\end{equation}
In both cases, $\mathcal{I}^2 = \mathbb{1}$. The two options are equivalent in the Hermitian case. For systems with a skin effect, pseudo-inversion is the relevant symmetry. With standard inversion symmetry,
\begin{equation} \label{eq: w1Dzero_with_standard_inversion}
w_{\mathrm{1D}} = \int_{0}^{2\pi} \frac{\mathrm{d}k}{2\pi \mathrm{i}} \frac{\mathrm{d}}{\mathrm{d} k} \mathrm{log}\,\mathrm{det}\,\mathcal{H}(-k) = -w_{\mathrm{1D}},
\end{equation}
so that the only admissible winding number is $w_{\mathrm{1D}} = 0$. And indeed, the Bloch Hamiltonian in Eq.~\eqref{eq: HatanoNelsonBloch} satisfies pseudo-inversion: $\mathcal{H}(k) = \mathcal{H}^\dagger(-k)$, so that $\mathcal{I} = 1$.

\begin{figure}[t]
    \includegraphics[width=\linewidth]{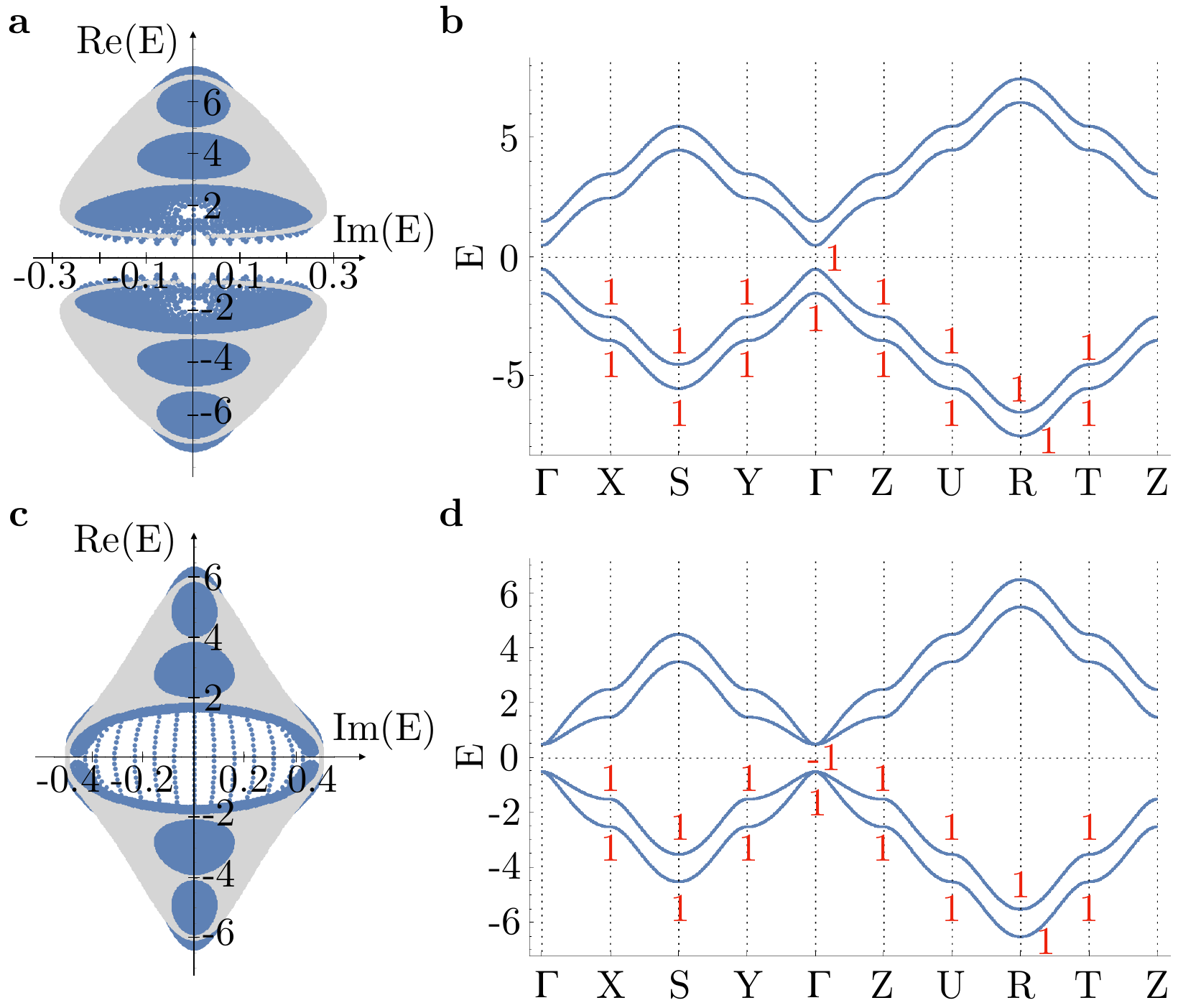}
    \caption{\label{fig: ETI_indicators}Inversion symmetry indicator for $w_{\mathrm{3D}}$. \textbf{a},~\textbf{c}~Bulk (grey) and surface (blue) spectra in the complex plane for a non-Hermitian topological insulator with $w_{\mathrm{3D}} = 0$ (\textbf{a}) and $w_{\mathrm{3D}} = 1$ (\textbf{c})~\cite{fn}. 
    \textbf{b},~\textbf{d}~Band structure of the corresponding Hermitian doubles along high-symmetry lines of the Brillouin zone. All bands are two-fold degenerate, and each pair has equal inversion eigenvalues at the inversion-symmetric momenta (the inversion eigenvalues of the occupied bands are shown in red). By Eq.~\eqref{eq: w3D_inversion_indicator}, the presence (absence) of a double band inversion in \textbf{d} (in \textbf{b}) indicates the presence (absence) of anomalous non-Hermitian surface states in \textbf{c} (in \textbf{a}).}
\end{figure}

Let us therefore consider a general 1D non-Hermitian Bloch Hamiltonian $\mathcal{H}(k)$ that satisfies Eq.~\eqref{eq: PseudoInversionConstraint}. Its Hermitian double is given by
\begin{equation} \label{eq: HermitianDoubleDefinition}
\bar{\mathcal{H}}(k) = \begin{pmatrix} 0 & \mathcal{H}(k) \\ \mathcal{H}^\dagger(k) & 0 \end{pmatrix}.
\end{equation}
The presence of a point gap of $\mathcal{H}(k)$ around $E=0$ translates into a gapped spectrum of $\bar{\mathcal{H}}(k)$.
By construction, $\bar{\mathcal{H}}(k)$ furthermore enjoys a chiral (sublattice) symmetry:
\begin{equation} \label{eq: chiral_sym_definition}
\bar{\mathcal{C}}\bar{\mathcal{H}}(k) \bar{\mathcal{C}}^\dagger = - \bar{\mathcal{H}}(k), \quad \bar{\mathcal{C}} = \begin{pmatrix} \mathbb{1} & 0 \\ 0 & - \mathbb{1} \end{pmatrix},
\end{equation}
positioning the Hermitian double in symmetry class AIII of the Altland-Zirnbauer classification [on the other hand, the non-Hermitian Hamiltonian $\mathcal{H}(k)$ has no internal symmetries and so lies in class A]. The quantity $w_{\mathrm{1D}}$ is then nothing but the winding number of 1D Hermitian systems with chiral symmetry, and as such counts the number of protected zero-energy edge states. Moreover, the pseudo-inversion symmetry of $\mathcal{H}(k)$ implies a standard inversion symmetry for $\bar{\mathcal{H}}(k)$: 
\begin{equation} \label{eq: HermitianDoubleInversion1D}
\bar{\mathcal{I}}\bar{\mathcal{H}}(k) \bar{\mathcal{I}}^\dagger = \bar{\mathcal{H}}(-k), \quad \bar{\mathcal{I}} = \begin{pmatrix} 0 & \mathcal{I} \\ \mathcal{I} & 0 \end{pmatrix}.
\end{equation}
(Note that $\bar{\mathcal{I}}$ and $\bar{\mathcal{C}}$ anti-commute if the non-Hermitian Hamiltonian has pseudo-inversion, whereas for standard inversion symmetry $\bar{\mathcal{I}}$ and $\bar{\mathcal{C}}$ commute and so preclude any symmetry-indicated band inversions.)
Correspondingly, the eigenstates of $\bar{\mathcal{H}}(k)$ can be chosen as eigenstates of $\bar{\mathcal{I}}$ at the inversion-symmetric momenta $k = 0,\pi$. It is well known~\cite{ChiuReview16,Neupert2018} that $w_{\mathrm{1D}}$ is related to the Zak phase $\gamma$ of $\bar{\mathcal{H}}(k)$ via 
\begin{equation} \label{eq: w1D_indicator_formula}
w_{\mathrm{1D}} \mod 2 = \gamma/\pi.
\end{equation}
Furthermore, Ref.~\onlinecite{Alexandradinata14} showed that the $\mathbb{Z}_2$-valued Zak phase is constrained by the relation
\begin{equation} \label{eq: w1D_indicator_formula2}
\gamma = \frac{\pi}{2} \sum_{k^* \in \mathcal{I}\mathrm{SMs}} e^{\mathrm{i} k^*} \mathrm{Tr}\left[\bar{\mathcal{I}} \bar{P}(k^*)\right] \mod 2\pi,
\end{equation}
where $\bar{P}(k)$ is the projector onto the occupied subspace of $\bar{\mathcal{H}}(k)$, and the inversion-symmetric momenta are given by $\mathcal{I}\mathrm{SMs} = \{0,\pi\}$. Eqs.~\eqref{eq: w1D_indicator_formula} and~\eqref{eq: w1D_indicator_formula2} provide the first example of a symmetry indicator formula for a non-Hermitian topological invariant.

\emph{3D winding number---}We next formulate symmetry indicator invariants for 3D point-gapped non-Hermitian Hamiltonians $\mathcal{H}(\bs{k})$ that are characterized by the 3D winding number~\cite{KawabataClassification19,UedaPaper18,Das:2019}
\begin{equation}
    \label{eq: UedaInvariant}
    w_{\mathrm{3D}} = - \sum_{ijk} \int_{\mathrm{BZ}} \frac{\mathrm{d}^3\bs{k}}{24 \pi^2}\epsilon_{ijk} \mathrm{Tr}\left[Q_{i}(\bs{k})Q_{j}(\bs{k})Q_{k}(\bs{k})\right] \in \mathbb{Z},
\end{equation}
where $Q_{j}(\bs{k}) = \mathcal{H}(\bs{k})^{-1} \partial_{k_{j}} \mathcal{H}(\bs{k})$, $j=x,y,z$, and $\mathrm{BZ}=[0,2\pi]^3$ denotes the 3D Brillouin zone. A nontrivial $w_{\mathrm{3D}}$ has been identified with the presence of anomalous non-Hermitian surface states in Ref~\onlinecite{denner2020exceptional}, giving rise to exceptional topological insulators (ETIs). Again, the Hermitian double $\bar{\mathcal{H}}(\bs{k})$, defined in analogy to Eq.~\eqref{eq: HermitianDoubleDefinition}, describes a gapped 3D system in Altland-Zirnbauer class AIII. The invariant $w_{\mathrm{3D}}$ then counts the number of protected (twofold) Dirac cone surface states~\cite{Schnyder08}.

To make contact with the 1D case, we begin by discussing systems with inversion symmetry. In analogy to Eq.~\eqref{eq: w1Dzero_with_standard_inversion}, standard inversion symmetry implies $w_{\mathrm{3D}} = 0$, prompting us to focus on pseudo-inversion [Eq.~\eqref{eq: PseudoInversionConstraint}].
Defining the inversion operator $\bar{\mathcal{I}}$ of the Hermitian double as in Eq.~\eqref{eq: HermitianDoubleInversion1D}, we now prove 
\begin{equation} \label{eq: w3D_inversion_indicator}
w_{\mathrm{3D}} \mod 2 = \nu_2, 
\end{equation}
where the $\mathbb{Z}_2$-valued symmetry indicator $\nu_2$ is defined by 
\begin{equation} \label{eq: w3D_inversion_indicator2}
\nu_2 = \frac{1}{4} \sum_{\bs{k}^* \in \mathcal{I}\mathrm{SMs}} \mathrm{Tr} \left[\bar{\mathcal{I}}\bar{P}(\bs{k}^*)\right] \mod 2.
\end{equation}
Here, $\bar{P}(\bs{k})$ is again the projector onto the occupied subspace of the Hermitian double $\bar{\mathcal{H}}(\bs{k})$, while the inversion-symmetric momenta are now drawn from 
\begin{equation}
\begin{aligned}
\mathcal{I}\mathrm{SMs} = \{&(0,0,0),(0,\pi,0),(\pi,0,0),(\pi,\pi,0),\\ &(0,0,\pi),(0,\pi,\pi),(\pi,0,\pi),(\pi,\pi,\pi)\}. 
\end{aligned}
\end{equation}
To derive the relation, we begin by noting that $\nu_2 = 1$ is precisely the condition for a gapped, inversion-symmetric Hermitian Hamiltonian (for now without chiral symmetry) to be a symmetry-indicated axion insulator (AXI)~\cite{AshvinAxion2,Varnava:2018aa,wieder2018axion}. AXIs are (time-reversal broken) higher-order topological insulators that host one-dimensional unidirectional (chiral) hinge states when cut into inversion-symmetric geometries. Their surfaces each host a gapped Dirac cone as a remnant of the bulk double band inversion (indicated by $\nu_2 = 1$). In fact, it is instructive to regard the hinge states of an AXI as domain wall bound states in the surface Dirac mass~\cite{HingeSM}. To prove Eq.~\eqref{eq: w3D_inversion_indicator} we then only need to show that the surfaces of an AXI respecting inversion \emph{and} chiral symmetry must remain gapless, because $w_{\mathrm{3D}}$ counts the number of protected surface Dirac cones of the Hermitian double. But this must be the case, because the unidirectional dispersion of chiral hinge states, whose presence is implied by any inversion-preserving surface gap, violates chiral symmetry as defined in Eq.~\eqref{eq: chiral_sym_definition}. 
(See Fig.~\ref{fig: ETI_indicators}.)

To motivate this point further, consider the Dirac surface theory of an AXI with chiral symmetry (where the surface normal is chosen to lie along the $z$-direction)~\cite{Khalaf17},
\begin{equation}
 \bar{\mathcal{H}}_\mathrm{D}(k_x, k_y) = k_x \sigma_x + k_y \sigma_y,
\end{equation}
where $\sigma_{x,y,z}$ is a set of Pauli matrices. Here, chiral symmetry is represented by $\bar{\mathcal{C}}_\mathrm{D} = \sigma_z$. The only mass term, multiplying $\sigma_z$, does not anti-commute with chiral symmetry and is therefore disallowed. Generalizing this, we conclude that all surfaces remain gapless and host a single Dirac cone, vindicating Eq.~\eqref{eq: w3D_inversion_indicator}.

It is possible to calculate the symmetry indicator $\nu_2$ without explicitly constructing the Hermitian double. For this, we utilize the singular-value decomposition~\footnote{Similarly, one may leverage the singular value decomposition to express Eq.~\eqref{eq: w1D_indicator_formula2} without reference to the Hermitian double.}, which was previously fruitfully applied in the context of non-Hermitian topological systems~\cite{HerviouSVD19}:
\begin{equation}
\mathcal{H}(\bs{k}) = U(\bs{k}) \Sigma(\bs{k}) V(\bs{k})^\dagger, 
\end{equation}
where $U(\bs{k})$, $V(\bs{k})$ are unitary matrices of the same dimension as $H(\bs{k})$, and $\Sigma(\bs{k})$ is a diagonal matrix with non-negative entries.
We have that
\begin{equation}
\mathcal{H}(\bs{k}) V(\bs{k}) = U(\bs{k}) \Sigma(\bs{k}), \quad \mathcal{H}(\bs{k})^\dagger U(\bs{k}) = V(\bs{k}) \Sigma(\bs{k})
\end{equation}
implying that the matrix
\begin{equation}
S(\bs{k}) = \frac{1}{\sqrt{2}} \begin{bmatrix}U(\bs{k}) & U(\bs{k}) \\V(\bs{k}) & -V(\bs{k}) \end{bmatrix}
\end{equation}
diagonalizes the Hermitian double $\bar{\mathcal{H}}(\bs{k})$:
\begin{equation}
S(\bs{k})^\dagger \bar{\mathcal{H}}(\bs{k}) S(\bs{k}) = 
\begin{bmatrix}
\Sigma(\bs{k}) & 0 \\ 0 & - \Sigma(\bs{k})
\end{bmatrix}.
\end{equation}
We therefore find
\begin{equation}
\begin{aligned}
\nu_2 = -\frac{1}{8} \sum_{\bs{k}^* \in \mathcal{I}\mathrm{SMs}} \mathrm{Tr} \big[&V(\bs{k})^\dagger \mathcal{I} U(\bs{k}) \\&+ U(\bs{k})^\dagger \mathcal{I} V(\bs{k})\big] \mod 2.
\end{aligned}
\end{equation}

\emph{Chiral Chern-Simons form---}We now turn to 3D non-Hermitian insulators in symmetry class AII. This class is characterized by a standard time-reversal symmetry, so that the non-Hermitian Bloch Hamiltonian and the Hermitian double satisfy
\begin{equation} \label{eq: StandardTRSConstraint}
\begin{aligned}
&\mathcal{T} \mathcal{H}(\bs{k})\mathcal{T}^\dagger = \mathcal{H}(-\bs{k}) \\
&\bar{\mathcal{T}}\bar{\mathcal{H}}(\bs{k}) \bar{\mathcal{T}}^\dagger = \bar{\mathcal{H}}(-\bs{k}), \quad \bar{\mathcal{T}} = \begin{pmatrix}\mathcal{T} & 0\\ 0& \mathcal{T} \end{pmatrix},
\end{aligned}
\end{equation}
for some anti-unitary operator $\mathcal{T}$. Together with the 3D version of Eq.~\eqref{eq: chiral_sym_definition}, we then have that time-reversal and chiral symmetry commute: $[\bar{\mathcal{T}},\bar{\mathcal{C}}] = 0$, locating the Hermitian double in Altland-Zirnbauer class CII.
As previously explained, Eq.~\eqref{eq: StandardTRSConstraint} necessitates $w_{\mathrm{3D}} = 0$. Nevertheless, the classification of non-Hermitian systems in Altland-Zirnbauer class AII (corresponding to a Hermitian double in class CII) is given by $\mathbb{Z}_2$~\cite{KawabataClassification19}, it is indicated by the chiral Chern-Simons form $\mathrm{CS}_3$ whose evaluation requires a smooth gauge of Bloch states over the Brillouin zone~\cite{ChiuReview16}. Note that the ETI with $w_{\mathrm{3D}} = 1$ is adiabatically related to a $k_z$-indexed pumping cycle of a 2D integer quantum Hall effect (with Chern number $1$) around the complex point gap~\cite{denner2020exceptional}. Then, appealing to the correspondence between the $\mathbb{Z}$-classified integer quantum Hall effect in class A and the $\mathbb{Z}_2$-classified quantum spin Hall effect in class AII, it is straightforward to construct a representative of a nontrivial point-gapped phase in class AII: we take two time-reversal related copies of the ETI and form the tensor sum
\begin{equation} \label{eq: TETI_ham}
\mathcal{H}(\bs{k}) = \begin{pmatrix} \mathcal{H}_{\mathrm{ETI}}(\bs{k}) & 0 \\ 0 & T^\dagger \mathcal{H}_{\mathrm{ETI}}(-\bs{k})T \end{pmatrix},
\end{equation}
where $T$ is an anti-unitary operator that squares to $-1$. The complex spectrum of $\mathcal{H}(\bs{k})$ can develop a real line gap under the addition of symmetry-allowed perturbations. However, we will see that the surface spectrum necessarily remains gapless (in that there is no line gap).
The non-Hermitian Hamiltonian satisfies Eq.~\eqref{eq: StandardTRSConstraint} with 
\begin{equation}
\mathcal{T} = \begin{pmatrix} 0 & T \\ T^\dagger & 0 \end{pmatrix}.
\end{equation}
Because time-reversal symmetry commutes with pseudo-inversion symmetry, the inversion eigenvalues of the Hermitian double form two copies of the inversion eigenvalues of an AXI and we obtain $\nu_4 = 2$, where $\nu_4$ is the symmetry indicator
\begin{equation} \label{eq: w3D_inversion_TRS_indicator2}
\nu_4 = \frac{1}{4} \sum_{\bs{k}^* \in \mathcal{I}\mathrm{SMs}} \mathrm{Tr} \left[\bar{\mathcal{I}}\bar{P}(\bs{k}^*)\right] \mod 4
\end{equation}
associated with the simultaneous presence of inversion and time-reversal symmetry.
Furthermore, we will show that the surfaces of $\mathcal{H}(\bs{k})$ are gapless and anomalous, in that they form two time-reversal related copies of the anomalous surface states of an ETI.
These observations prompt us to posit
\begin{equation} \label{eq: CS_symmetryIndicator}
\mathrm{CS}_3 \mod 2 = \frac{1}{2}\nu_4 \mod 2,
\end{equation}
where 
\begin{equation} \label{eq: ChernSimonsInvariant}
\begin{aligned}
\mathrm{CS}_3 = - \sum_{ijk} \int_{\mathrm{BZ}} \frac{\mathrm{d}^3\bs{k}}{8 \pi^2} \epsilon_{ijk} \mathrm{Tr} \bigg[&A_i(\bs{k}) \partial_{k_j} A_k(\bs{k}) \\&+ \frac{2}{3} A_i(\bs{k}) A_j(\bs{k}) A_k(\bs{k}) \bigg],
\end{aligned}
\end{equation}
is the chiral Chern-Simons form, the invariant for Hermitian insulators in class CII~\cite{KawabataClassification19,ChiuReview16}. A nontrivial $\mathrm{CS}_3$ corresponds to a stable pair of Dirac cones in the surface spectrum of the Hermitian double~\cite{Schnyder08}, whose presence is heralded by $\nu_4 = 2$ (recall that we have previously shown that the surfaces of an AXI host a single Dirac cone as long as chiral symmetry is enforced). Here, we have made use of the non-Abelian, matrix-valued Berry connection
\begin{equation}
[A_i(\bs{k})]^{mn} = \braket{u^m (\bs{k}) |\partial_{k_i}| u^n (\bs{k})},
\end{equation}
that is defined in terms of the occupied (negative-energy) Bloch eigenstates $\ket{u^n (\bs{k})}$ of $\bar{\mathcal{H}}(\bs{k})$. We note that Eq.~\eqref{eq: ChernSimonsInvariant} is gauge dependent and needs to be supplemented by the gauge condition
\begin{equation} \label{eq: CSgaugeConditions}
\int_{\partial\mathrm{BZ}_{1/2}} \mathrm{d}^2\bs{k} \epsilon_{ij} \mathrm{Tr} \left\{[X(\bs{k}) \partial_{k_i} X^\dagger(\bs{k})] [X(\bs{k}) \partial_{k_j} X^\dagger(\bs{k})]\right\} = 0,
\end{equation}
where 
\begin{equation}
X(\bs{k}) = [\ket{u^1 (\bs{k})}, \dots, \ket{u^N (\bs{k})}, \ket{v^1 (\bs{k})}, \dots, \ket{v^N (\bs{k})}]
\end{equation}
is the unitary matrix of occupied and unoccupied eigenstates of $\bar{\mathcal{H}}(\bs{k})$, respectively. Evidently, the integral in Eq.~\eqref{eq: ChernSimonsInvariant} and its associated gauge condition in Eq.~\eqref{eq: CSgaugeConditions} are sufficiently involved as to make an expression in terms of symmetry indicators highly desirable.

For Eq.~\eqref{eq: CS_symmetryIndicator} to be well-defined, $\nu_4 = 1,3$ should not be admissible. And indeed, by the Fu-Kane criterion~\cite{FuKane07}, these values indicate an odd number of double band inversions in the bulk of the Hermitian double [$2(2n+1)$, $n \in \mathbb{Z}$, inversion eigenvalues of the valence manifold switch sign with respect to the atomic limit], however, such a scenario is disallowed in symmetry class CII~\cite{Schnyder08}.

Finally, we discuss the surface physics of $\mathcal{H}(\bs{k})$ in Eq.~\eqref{eq: TETI_ham}. In the unperturbed case, Eq.~\eqref{eq: TETI_ham}, each ETI contributes a single exceptional point to the complex surface spectrum. We can model the surface with $z$-normal by the Dirac Hamiltonian
\begin{equation}
    \mathcal{H}_\mathrm{D}(k_x, k_y) = (k_x+\mathrm{i} c_x) \sigma_x + (k_y + \mathrm{i} c_y) \sigma_y + \mathrm{i} c_z \sigma_z,
\end{equation}
where $c_{x,y,z}$ are real numbers and $\sigma_{x,y,z}$ is a set of Pauli matrices. This Hamiltonian satisfies standard time-reversal symmetry with $\mathcal{T} = \mathrm{i} \sigma_y \mathit{K}$, where $\mathit{K}$ denotes complex conjugation. For generic values of $c_{x,y,z}$, there are two exceptional points in the spectrum of $\mathcal{H}_\mathrm{D}(k)$. These may morph into an exceptional loop for $c_x = c_y = 0$, or into a Hermitian Dirac cone for $c_x = c_y = c_z = 0$, but since $\mathcal{H}_\mathrm{D}(k)$ exhausts all terms allowed by symmetry to linear order in $k$ (up to unitary basis transformations and terms multiplying the identity matrix), they can never annihilate. We therefore conclude that the surface must host two exceptional points. In the Supplemental Material, we discuss this case and the pseudo time-reversal symmetric case with $\nu_4 = 2$. There, we also present tight-binding models for all phases discussed here.

\emph{Discussion---}
We have derived symmetry indicator invariants for $w_{\mathrm{1D}}$ and $w_{\mathrm{3D}}$, which classify point-gapped non-Hermitian insulators in Altland-Zirnbauer symmetry class A. Furthermore, we showed that non-Hermitian time-reversal symmetric topological insulators can be identified using symmetry indicators without the need for a smooth Bloch gauge. In order to connect non-Hermitian topological invariants with the symmetry indicators of the Hermitian double, our strategy was to study the effect of chiral symmetry on Hermitian topological crystalline insulators. Interestingly, the introduction of chiral symmetry changes the bulk-boundary correspondence of these insulators, while leaving their symmetry indicators intact. This approach can be straightforwardly generalized to other symmetry classes.

\emph{Note added---}
While preparing this manuscript, we became aware of a recent related work~\cite{Okugawa2021}, which also investigates non-Hermitian symmetry indicators, but focuses on bulk exceptional points and lines in 2D and 3D.

\bibliography{Ref-Lib}

\end{document}


\title{Supplementary Material for\\ ``Symmetry Indicators for Non-Hermitian Topological Band Structures"}

\author{Pascal M. Vecsei}
\affiliation{Department of Physics, University of Zurich, Winterthurerstrasse 190, 8057 Zurich, Switzerland}

\author{M. Michael Denner}
\affiliation{Department of Physics, University of Zurich, Winterthurerstrasse 190, 8057 Zurich, Switzerland}

\author{Titus Neupert}
\affiliation{Department of Physics, University of Zurich, Winterthurerstrasse 190, 8057 Zurich, Switzerland}

\author{Frank Schindler}
\affiliation{Princeton Center for Theoretical Science, Princeton University, Princeton, NJ 08544, USA}

\maketitle

\tableofcontents

\section{Models}
In this section, tight-binding models that allow for the application of the symmetry indicator invariants presented in the main text, as well as models with an additional (pseudo) time reversal symmetry, are listed.
\subsection{1D model for $\gamma = \pi$, $w_\text{1D}=1$}
A 1D model with pseudo inversion symmetry is given by
\begin{equation}
    H(k) = t_r e^{ik} + t_l e^{-ik},
\end{equation}
with $\Inv = 1$ and $w_\text{1D} = \text{sign}(t_r - t_l)$. Its Hermitian double Hamiltonian is given by
\begin{equation}
    \bar H(k) = \begin{pmatrix}0 & t_r e^{ik} + t_l e^{-ik} \\ t_r e^{-ik} + t_l e^{ik} & 0\end{pmatrix}
\end{equation}
and has inversion symmetry represented by $\bar \Inv = \sigma_x$. For $t_r > t_l$, the Hamiltonian is in the topologically nontrivial phase with $w_\text{1D} = +1$. The inversion eigenvalues at the inversion-symmetric momenta in the occupied bands of the Hermitian double are then
\begin{equation}
\begin{split}
    k &= 0 \to -1,\\
    k &= \pi \to 1,\\
\end{split}
\end{equation}
therefore $\gamma = \pi$ and $w_\text{1D}$ has to be odd.

\subsection{Model for $\nu_2 = 1$, $w_\text{3D} = 1$}
In order to study a system with $w_\text{3D} = +1$ and pseudo-inversion symmetry, we consider the model for an exceptional topological insulator (ETI)~\cite{denner2020exceptional}. Its Hamiltonian is given by
\begin{equation}
\begin{split}
    H(\mathbf k) =& \left( \sum_{j = x,y,z} \cos(k_j) - M \right) \tau_z \sigma_0 \\&+ \lambda \sum_{j=x,y,z} \sin(k_j) \tau_x \sigma_j + i \delta \tau_x \sigma_0,
\end{split}
\end{equation}
where the Pauli matrices $\sigma_{\mu}$ and $\tau_{\mu}$ act on the spin and orbital degrees of freedom, respectively, with $\mu = 0, x, y, z$ and the 0-th Pauli matrix as the $2\times2$ identity matrix. 
The invariant is $w_\text{3D} = 1$ for $3 - \delta/\lambda < M < 3 + \delta/\lambda$. By changing $M$, we transition to the trivial phase at $M=3 \pm \delta/\lambda$. The Hamiltonian has pseudo-inversion symmetry with $\Inv = \tau_z \sigma_0$ and pseudo time-reversal symmetry represented by $\Trs = \tau_0 \sigma_y \mathit{K}$, where $\mathit{K}$ denotes complex conjugation.
Correspondingly, the Hermitian double Hamiltonian takes the form
\begin{equation}
\begin{split}
    \bar H (\mathbf k) =& \left( \sum_{j = x,y,z} \cos(k_j) - M \right) \eta_x \tau_z \sigma_0 \\&+ \lambda \sum_{j=x,y,z} \eta_x  \sin(k_j) \tau_x \sigma_j - \eta_y \delta \tau_x \sigma_0,
\end{split}
\end{equation}
and has time-reversal and inversion symmetry, represented by
\begin{equation}
    \bar \Trs = \eta_x \tau_0 \sigma_y \mathit{K} \hspace{1mm} \text{ and } \hspace{1mm} \bar \Inv = \eta_x \tau_z \sigma_0,
\end{equation}
respectively. $\eta_\mu$ are an additional set of Pauli matrices. The inversion symmetry eigenvalues in the occupied bands of the Hermitian double Hamiltonian are given by
\begin{equation}
\begin{split}
    \bs{k} &= (0,0,0) \to (1,1,-1,-1),\\
    \bs{k} &= (0,0,\pi) \to (1,1,1,1),\\
    \bs{k} &= (0,\pi,0) \to (1,1,1,1),\\
    \bs{k} &= (0,\pi,\pi) \to (1,1,1,1),\\
    \bs{k} &= (\pi,0,0) \to (1,1,1,1),\\
    \bs{k} &= (\pi,0,\pi) \to (1,1,1,1),\\
    \bs{k} &= (\pi,\pi,0) \to (1,1,1,1),\\
    \bs{k} &= (\pi,\pi,\pi) \to (1,1,1,1).\\
\end{split}
\end{equation}
Therefore we find $\nu_2 = 1$.
\subsection{Pseudo time-reversal-symmetric models for $\nu_4 = 2$, $w_\text{3D} = 0$ and $w_\text{3D} = 2$}

\begin{figure*}[t]
\centering
\includegraphics[width=\textwidth]{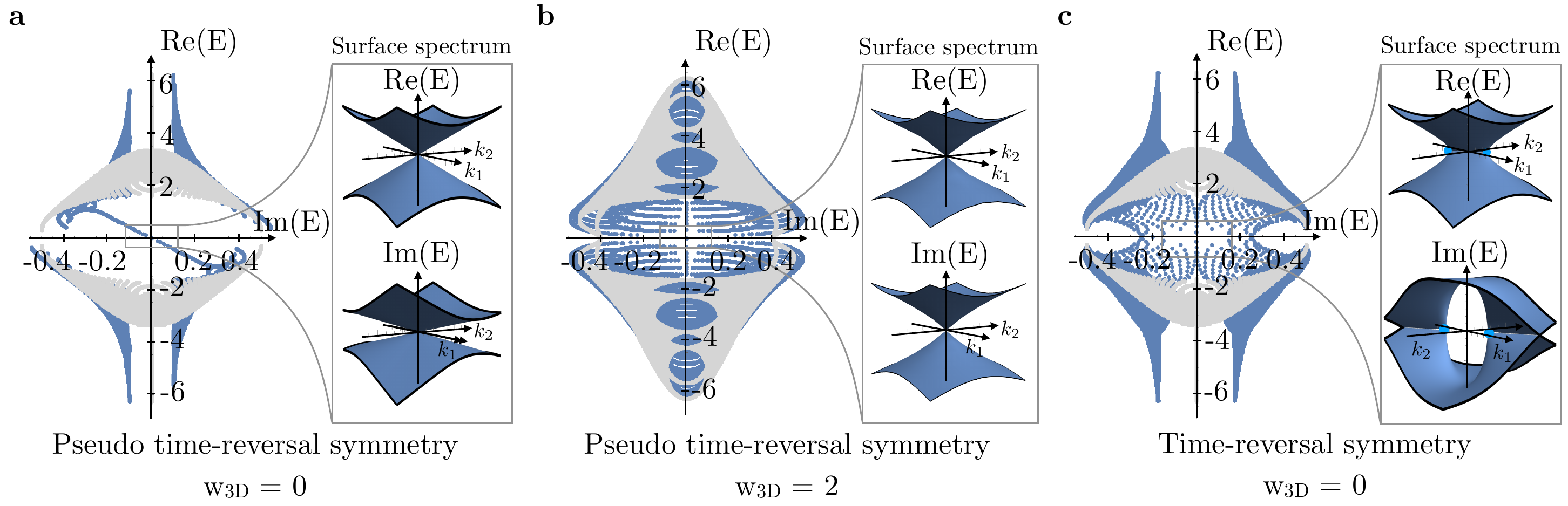}
\caption{\label{fig:sym_ETIs} Bulk and surface spectra of (pseudo) time-reversal symmetric ETIs. \textbf{a} The surface state filling the line gapped bulk (grey) of the TRS$^\dag$ ETI shows a complex Dirac cone in the surface spectrum (blue) (We use \eqref{eq:PseudoTRS-ETI} with $M = 3, \lambda = 1, \delta = 0.5$ and $\gamma = 0.1$). \textbf{b} The point gapped pseudo time-reversal symmetric ETI with $w_{\text{3D}} = 2$ (grey) shows a similar surface dispersion, with the states filling the point gap (blue) forming again a complex Dirac cone [We use \eqref{eq:PseudoTRS-ETI-w3D2} with $M = 3, \lambda = 1, \delta = 0.5$ and $\gamma = 0.05$]. \textbf{c} The bulk spectrum of the TRS ETI shows a line gap (grey) which is filled with surface states (blue), forming a pair of exceptional points in the surface Brillouin zone (We use \eqref{eq:TRS-ETI} with $M = 3, \lambda = 1, \delta = 0.5$ and $|\gamma| = 0.1$).}
\end{figure*}

In order to study the surface spectrum of the pseudo-time-reversal symmetric ETI shown in Fig.~\ref{fig:sym_ETIs}~(b), we consider copies of the ETI with opposite $w_\text{3D} = \pm 1$,

\begin{equation}
\begin{split}
    H(\bs{k})&=\left(\sum_{j=x,y,z} \cos k_j-M\right) \rho_0\tau_z\sigma_0\\
    &\quad+\lambda  \sum_{j=x,y,z}\sin k_j \, \rho_0\tau_x\sigma_j+\mathrm{i}\delta\,\rho_z\tau_x\sigma_0 + \mathrm{i}\gamma \,\rho_x\tau_z\sigma_0,
    \label{eq:PseudoTRS-ETI}
\end{split}
\end{equation}
which possesses pseudo time-reversal symmetry as $\Trs = \rho_0 \tau_0 \sigma_y \mathit{K}$ and pseudo-inversion symmetry with $\Inv = \rho_x \tau_z \sigma_0$. Again, the transition from the topologically nontrivial to the trivial phase can be achieved by tuning $M$. Its bulk (open and periodic boundary conditions) and surface spectra are shown in  Fig.~\ref{fig:sym_ETIs}~(a). The Hermitian double Hamiltonian is then given by 
\begin{equation}
\begin{split}
    \bar H(\bs{k})&=\left(\sum_{j=x,y,z} \cos k_j-M\right) \eta_x \rho_0\tau_z\sigma_0\\
    &\quad+\lambda  \sum_{j=x,y,z}\sin k_j \, \eta_x \rho_0\tau_x\sigma_j \\&\quad- \delta\, \eta_y \rho_z\tau_x\sigma_0 - \gamma \, \eta_y \rho_x\tau_z\sigma_0,
\end{split}
\end{equation}
and carries pseudo time-reversal and inversion symmetry, represented by
\begin{equation}
    \bar \Trs = \eta_x \rho_0 \tau_0 \sigma_y \mathit{K} \hspace{1mm} \text{ and } \hspace{1mm} \bar \Inv = \eta_x \rho_x \tau_z \sigma_0,
\end{equation}
respectively. The inversion symmetry eigenvalues in the occupied bands of the Hermitian double Hamiltonian are given by
\begin{equation}
\begin{split}
    \bs{k} &= (0,0,0) \to (1,1,1,1,-1,-1,-1,-1),\\
    \bs{k} &= (0,0,\pi) \to (1,1,1,1,1,1,1,1),\\
    \bs{k} &= (0,\pi,0) \to (1,1,1,1,1,1,1,1),\\
    \bs{k} &= (0,\pi,\pi) \to (1,1,1,1,1,1,1,1),\\
    \bs{k} &= (\pi,0,0) \to (1,1,1,1,1,1,1,1),\\
    \bs{k} &= (\pi,0,\pi) \to (1,1,1,1,1,1,1,1),\\
    \bs{k} &= (\pi,\pi,0) \to (1,1,1,1,1,1,1,1),\\
    \bs{k} &= (\pi,\pi,\pi) \to (1,1,1,1,1,1,1,1).\\
\end{split}
\end{equation}
Therefore we find $\nu_4 = 2$.

Similarly we can construct a Hamiltonian with winding number $w_{\text{3D}} = 2$ by combining two ETIs with $w_\text{3D} = +1$,
\begin{equation}
\begin{split}
    H(\bs{k})&=\left(\sum_{j=x,y,z} \cos k_j-M\right) \rho_0\tau_z\sigma_0\\
    &\quad+\lambda  \sum_{j=x,y,z}\sin k_j \, \rho_0\tau_x\sigma_j+\mathrm{i}\delta\,\rho_0\tau_x\sigma_0 + \mathrm{i}\gamma \,\rho_x\tau_x\sigma_0,
    \label{eq:PseudoTRS-ETI-w3D2}
\end{split}
\end{equation}
now possessing a pseudo-inversion symmetry $\Inv = \rho_0 \tau_z \sigma_0$ and pseudo time-reversal symmetry with $\Trs = \rho_0 \tau_0 \sigma_y \mathit{K}$. The bulk and surface spectrum of this model is shown in Fig.~\ref{fig:sym_ETIs}~(b). Its Hermitian double Hamiltonian is given by
\begin{equation}
\begin{split}
    \bar H(\bs{k})&=\left(\sum_{j=x,y,z} \cos k_j-M\right) \eta_x \rho_0\tau_z\sigma_0\\
    &\quad+\lambda  \sum_{j=x,y,z}\sin k_j \, \eta_x \rho_0\tau_x\sigma_j\\
    &\quad-\delta\, \eta_y \rho_0\tau_x\sigma_0 -\gamma \,  \eta_y \rho_x\tau_x\sigma_0,
\end{split}
\end{equation}
and has inversion symmetry with $\bar \Inv = \eta_x \rho_0 \tau_z \sigma_0$ and time-reversal symmetry with $\bar \Trs = \eta_x \rho_0 \tau_0 \sigma_y \mathit{K}$. The inversion eigenvalues in the occupied bands of the Hermitian double Hamiltonian are 
\begin{equation}
\begin{split}
    \bs{k} &= (0,0,0) \to (1,1,1,1,-1,-1,-1,-1),\\
    \bs{k} &= (0,0,\pi) \to (1,1,1,1,1,1,1,1),\\
    \bs{k} &= (0,\pi,0) \to (1,1,1,1,1,1,1,1),\\
    \bs{k} &= (0,\pi,\pi) \to (1,1,1,1,1,1,1,1),\\
    \bs{k} &= (\pi,0,0) \to (1,1,1,1,1,1,1,1),\\
    \bs{k} &= (\pi,0,\pi) \to (1,1,1,1,1,1,1,1),\\
    \bs{k} &= (\pi,\pi,0) \to (1,1,1,1,1,1,1,1),\\
    \bs{k} &= (\pi,\pi,\pi) \to (1,1,1,1,1,1,1,1).\\
\end{split}
\end{equation}
Therefore we find $\nu_4 = 2$.

\subsection{Time-reversal symmetric model for $\nu_4 = 2$}

A time-reversal symmetric ETI is formed by stacking two TRS related ETIs as

\begin{equation}
\begin{split}
    H(\bs{k})&=\left(\sum_{j=x,y,z} \cos k_j-M\right) \rho_0\tau_z\sigma_0\\
    &\quad+\lambda  \sum_{j=x,y,z}\sin k_j \, \rho_0\tau_x\sigma_j+\mathrm{i}\delta\,\rho_z\tau_x\sigma_0 + \mathrm{i} \,\rho_0\tau_z \boldsymbol{\gamma}\cdot\sigma,
    \label{eq:TRS-ETI}
\end{split}
\end{equation}
which possesses time-reversal symmetry as $\Trs = \rho_x \tau_0 \sigma_y \mathit{K}$ and pseudo-inversion symmetry with $\Inv = \rho_0 \tau_z \sigma_0$. The resulting energy spectra shown in Fig.~\ref{fig:sym_ETIs}~(c) highlight the emergence of two surface exceptional points, if the bulk spectrum exhibits a real line gap. Its Hermitian double Hamiltonian is 
\begin{equation}
\begin{split}
    H(\bs{k})&=\left(\sum_{j=x,y,z} \cos k_j-M\right) \eta_x \rho_0\tau_z\sigma_0\\
    &\quad+\lambda  \sum_{j=x,y,z}\sin k_j \, \eta_x \rho_0\tau_x\sigma_j\\
    &\quad-\delta\,  \eta_y \rho_z\tau_x\sigma_0 - \, \eta_y \rho_0\tau_z \boldsymbol{\gamma}\cdot\sigma,
\end{split}
\end{equation}
and has time-reversal symmetry with $\bar \Trs = \eta_0 \rho_x \tau_0 \sigma_y \mathit{K}$ and inversion symmetry with $\bar \Inv = \eta_x \rho_0 \tau_z \sigma_0$. The inversion eigenvalues in the occupied bands of the Hermitian double Hamiltonian are 
\begin{equation}
\begin{split}
    \bs{k} &= (0,0,0) \to (1,1,1,1,-1,-1,-1,-1),\\
    \bs{k} &= (0,0,\pi) \to (1,1,1,1,1,1,1,1),\\
    \bs{k} &= (0,\pi,0) \to (1,1,1,1,1,1,1,1),\\
    \bs{k} &= (0,\pi,\pi) \to (1,1,1,1,1,1,1,1),\\
    \bs{k} &= (\pi,0,0) \to (1,1,1,1,1,1,1,1),\\
    \bs{k} &= (\pi,0,\pi) \to (1,1,1,1,1,1,1,1),\\
    \bs{k} &= (\pi,\pi,0) \to (1,1,1,1,1,1,1,1),\\
    \bs{k} &= (\pi,\pi,\pi) \to (1,1,1,1,1,1,1,1).\\
\end{split}
\end{equation}
Therefore we find $\nu_4 = 2$.

\section{Exceptional topological insulators with time-reversal symmetry}
We study systems with (pseudo) time-reversal symmetry. 3D systems with pseudo time reversal symmetry (TRS$^\dag$) have a $\mathds{Z}$ classification with $w_\text{3D}$ as invariant. 3D systems with TRS have a $\mathds{Z}_2$ classification with the Chern-Simons invariant whose evaluation requires a specific gauge (cf. main text). In the following, we study systems with $\nu_4 = 2$, because $\nu_4$ is a $\mathds{Z}_4$ invariant in presence of pseudo time-reversal symmetry, and $\nu_4 = 2$ is the only case that is not already captured by the invariant $\nu_2$.
\subsection{Pseudo time-reversal symmetry}
If $\nu_4 = 2$ in a pseudo time-reversal symmetric systems, this can correspond to different topological phases:
\begin{enumerate}
    \item The Hamiltonian corresponds to a phase with $w_\text{3D} \mod 4 = 2$. This phase does not require crystal symmetries for its protection.
    \item The Hamiltonian has $w_\text{3D} \mod 4 = 0$. This can appear, for example, by combining two Hamiltonians with equal but opposite $w_\text{3D} = \pm 1$.
\end{enumerate}
Both of these phases have the same symmetry indicator, and both are topologically nontrivial, but if $w_\text{3D} = 0$, the bulk spectrum can obtain a real line gap by addition of perturbations compatible with pseudo-inversion and pseudo time-reversal symmetries. This phase in the bulk as well as on the surface, is depicted in Fig.~\ref{fig:sym_ETIs}~(a). Its spectrum shows a complex Dirac cone on the surface. It can be understood as the symmetry-respecting non-Hermitian deformation of a Hermitian strong topological insulator (TI), since after we have added a perturbation that opens a real line gap, we could further deform the Hamiltonian, while respecting the symmetries, until arriving at a Hermitian phase. This Hermitian phase still has the same symmetry indicator, $\nu_4 = 2$, but, using that $\text{Tr}(\Inv) = 0$, this implies that the symmetry indicator invariant of the deformed Hermitian Hamiltonian, calculated using the eigenvalues of the Hamiltonian itself, and not its Hermitian double, is
\begin{equation}\kappa_1 \mod 2 = \frac{1}{4} \sum_{\bs{k^*} \in \Inv\text{SMs}} \text{Tr}\left[ \Inv P(\bs{k^*})\right] \mod 2= 1.\end{equation}
Therefore, the deformed Hermitian Hamiltonian is a strong TI~\cite{Khalaf17}.

On the other hand, if $w_\text{3D}= 2$, no real line gap can be induced into the bulk spectrum by adding a symmetric perturbation. The bulk spectrum with open and periodic boundary conditions as well as the surface spectrum are depicted in Fig.~\ref{fig:sym_ETIs}~(b).

In general, if pseudo time-reversal symmetry is present, we can study the surface by studying its spectrum in a linearized regime. If $\Trs_\mathrm{s} = i \sigma_y \mathit{K}$ is the representation of pseudo time-reversal symmetry on the surface, the only allowed perturbation term is $c \sigma_0$ with $c \in \mathds{C}$ arbitrary. This constant perturbation cannot induce a point or line gap in the surface spectrum, because it only moves all the eigenvalues by the constant complex number $c$.

\subsubsection{Surface of the Hermitian double Hamiltonian}
A TRS$^\dag$ symmetry of the Hamiltonian implies a TRS of the Hermitian double Hamiltonian, which appears in addition to the already existing inversion and chiral symmetries.
We describe an argument using a surface Hamiltonian for the Hermitian double. We want to show that a surface Hamiltonian with two gapless states (a double Dirac cone) cannot be gapped out. If the surface of a system is in addition TR-symmetric, we have, for example,
\begin{equation}
 \Ham_\text{kin}(k_x, k_y) = k_x \rho_z \tau_z \sigma_x + k_y \rho_z \tau_z \sigma_y, 
\end{equation}
with the symmetries
\begin{equation}
 \begin{split}
  \Trs &= \rho_0 \tau_0 \sigma_y \mathit{K}, \\
  \Inv &= \rho_0 \tau_x \sigma_0, \\
  \Sls &= \rho_0 \tau_y \sigma_0.
 \end{split}
\end{equation}
In this situation a potential mass term has to commute with $\Trs$ and $\Inv$ and anticommute with $\Sls$ and the kinetic part. The only terms fulfilling the symmetry requirements are
\begin{equation}
\begin{split}
  \Ham_\text{mass}(k_x, k_y) &= m \rho_0 \tau_x \sigma_0,\\
  \Ham_\text{mass}(k_x, k_y) &= m \rho_z \tau_x \sigma_0,\\
  \Ham_\text{mass}(k_x, k_y) &= m \rho_y \tau_x \sigma_z.
\end{split}
\end{equation}
These terms are non-local, so they cannot serve as mass terms. The fact that a mass term is forbidden shows that there is indeed a $\mathds{Z}_4$ symmetry indicator, which unfortunately does not fix the topology of the system completely.

\subsection{Time-reversal symmetry}
If time-reversal symmetry is present, and the symmetry indicator is $\nu_4 = 2$, the surface spectra has two EPs, which can lie atop of each other. The bulk spectra with PBC and OBC, as well as the surface spectrum, are depicted in Fig.~\ref{fig:sym_ETIs}~(c). In this phase, symmetry respecting perturbations can also open a real line gap with periodic boundary conditions. This phase is, like the $w_\text{3D}=0$-phase with TRS$^\dag$, a non-Hermitian deformation of a Hermitian strong TI phase.

To show that no line gap can appear in the complex surface spectrum, a linearized regime on the surface is taken into consideration, with TRS represented as $i \sigma_y$.
We expect that the surface states are the time-reversal doubled states already known from the ETI~\cite{denner2020exceptional}. That would mean that either a $k_x + ik_y$ dispersion is doubled, or a system with two exceptional points (EPs) appears. We study two-band systems in a linearized regime. Our surface Hamiltonian is characterised by the 12 real parameters $\alpha^\mu= (\alpha^0 , \vec \alpha)$, $\beta^\mu = (\beta^0, \vec \beta)$ and $\gamma^\mu = (\gamma^0 , \vec \gamma)$, which act as prefactors to the allowed kinetic terms $\sigma^\mathrm{kin}_\mu = (\sigma^\mathrm{kin}_0, \vec \sigma^\mathrm{kin})$ and perturbation terms $\sigma^\mathrm{pert}_\mu = (\sigma^\mathrm{pert}_0 , \vec \sigma^\mathrm{pert})$. The allowed terms are
\begin{equation}
\begin{split}
  \sigma^\mathrm{pert}_0 &= \sigma_0,\\
  \sigma^\mathrm{pert}_1 &= i \sigma_x,\\
  \sigma^\mathrm{pert}_2 &= i \sigma_y,\\
  \sigma^\mathrm{pert}_3 &= i \sigma_z,\\
\end{split}
\end{equation}
and
\begin{equation}
\begin{split}
\sigma^\mathrm{kin}_0 &= i \sigma_0,\\
\sigma^\mathrm{kin}_1 &= \sigma_x,\\
\sigma^\mathrm{kin}_2 &= \sigma_y,\\
\sigma^\mathrm{kin}_3 &= \sigma_z .
\end{split}
\end{equation}
The two-band Dirac Hamiltonian is then, in full generality, given by
\begin{equation}
\label{twobanddiracHaminFullGenerality}
 \Ham(k_x, k_y) = k_x \alpha^\mu \sigma^\mathrm{kin}_\mu +k_y \beta^\mu \sigma^\mathrm{kin}_\mu +  \gamma^\mu \sigma^\mathrm{pert}_\mu,
\end{equation}
with summation over repeated indices. The question to be treated is whether it is possible to induce a gap into the spectrum on the surface while respecting TRS.
\subsubsection{$k_x + ik_y$ dispersion}
We start by studying the time-reversal symmetric doubling of the $k_x + i k_y$ dispersion of the ETI. The kinetic part of the Hamiltonian is given by
\begin{equation}
 \Ham_\text{kin}(k_x, k_y) = k_x \sigma_x + i k_y \sigma_0,
\end{equation}
to which an arbitrary perturbation term of form
\begin{equation}
 \Ham_\text{pert}(k_x, k_y) = \gamma^\mu \sigma_\mu^\text{pert}
\end{equation}
is added. The eigenvalues of this Hamiltonian are given by the equation
\begin{multline}
 0 = \gamma^\mu \gamma_\mu -k_x^2 - k_y^2 - 2\gamma^0 z + z^2 \\+ 2i (\gamma^0 k_y - \gamma^1 k_x - k_y z).
\end{multline}
Setting $z = x + iy$ with $x,y \in \mathds{R}$, we obtain 
\begin{multline}
 k_x^2 =  \vec \gamma^2 + (\gamma^0 - x)^2 - (k_y - y)^2 \\ \text{ and } -\gamma^1 k_x +(\gamma^0-x) (k_y -  y)= 0 .
\end{multline}
To check for which values of $(x,y)$ there is a solution $(k_x,k_y)$, we first look at the case $(\gamma^0 -x) \neq 0$. Then we have
\begin{multline}
 k_y - y = \frac{\gamma^1 k_x}{\gamma^0 -x} \\
 \implies \left(1+ \left(\frac{\gamma^1 }{\gamma^0 -x} \right)^2 \right)  k_x^2= \vec \gamma^2 + (\gamma^0 -x)^2 .
\end{multline}
Since $\left(1+ \left(\frac{\gamma^1 }{\gamma^0 -x} \right)^2 \right) \neq 0$, the equations yield two real solutions $k_x$, each of which permits a calculation of $k_y$. On the other hand, if $\gamma^0 -x = 0$, we get the two equations
\begin{equation}
 k_x^2 = \vec \gamma^2 - (k_y - y)^2 \text{ and } -\gamma^1 k_x = 0.
\end{equation}
We choose $k_x = 0$ in order to satisfy the second equation, and get
\begin{equation}
 (k_y -y)^2 = \vec \gamma^2.
\end{equation}
This equation can be solved for $k_y$ for any choice of $y$. Therefore, we get the result $(k_x, k_y) = (0, y \pm \abs{\vec \gamma})$ on the line with real part $x = \gamma^0$.

Thus, we have shown that for every point on the complex plane we can find a momentum $(k_x,k_y)$ such that the Hamiltonian has the complex number $E = x + i y$ as its eigenvalue. Thus, there is no gap.

\subsubsection{Two Exceptional Points}
The case of two exceptional Points (EPs) can be modelled with the 2-band Hamiltonian 
\begin{equation}
\label{ComplexTwoEPsHamiltonianFormula}
 \Ham(k_x, k_y) = k_x \sigma_x + k_y \sigma_z + i \sigma_x.
\end{equation}
This is an interesting situation because we know from Ref.~\onlinecite{denner2020exceptional} that one of the possible surface states of an ETI is a state with a single EP. Physically, one of the possible expected states of a time-reversal doubled ETI is therefore a state with two EPs.\\

Expressed as a generic Hamiltonian, the parameters are $\alpha^1 = 1$, $\beta^3 = 1$, $\gamma^1 = 1$ [cf. Eq.~\eqref{twobanddiracHaminFullGenerality}]. For simplicity, we omit $i\sigma_x$, which we are allowed to do since it is anyway a symmetry-allowed term, and below we allow for arbitary perturbations including this one. Now, we check whether $k_x \sigma_x + k_y \sigma_z$ can be gapped out by adding an arbitrary perturbation. The equations for $k_x, k_y$, given the eigenvalue $E = x + iy$, are
\begin{equation}
 \begin{split}
 \gamma_\mu \gamma^\mu  - k_x^2  - k_y^2 - 2 \gamma^0 x + x^2  - y^2 &= 0,\\
 -  \gamma^1 k_x-  \gamma^3 k_y-  \gamma^0 y +  x y &= 0.\\
 \end{split}
\end{equation}
As we know that $\gamma^1 \approx 1$, we can solve the second equation for $k_x$, which is 
\begin{equation}
 k_x = \frac{1}{\gamma^1} \left( x y - \gamma^0 y - \gamma^3 k_y \right).
\end{equation}

This can then be used to calculate $k_y$, given as
\begin{equation}
 k_y = \pm \frac{\sqrt{\vec \gamma \cdot \vec \gamma + ( \gamma^0 -x)^2}}{\sqrt{1 + \frac{(\gamma^3)^2}{(\gamma^1)^2}}}
\end{equation}
if $y=0$, which always yields a solution. If $y \neq 0$, we get
\begin{equation}
\begin{split}
 k_y &= \frac{\gamma^3 x - \gamma^0 \gamma^3}{(\gamma^1)^2 + (\gamma^3)^2} y \pm \frac{(\gamma^1)}{(\gamma^1)^2 + (\gamma^3)^2} \sqrt{\mathfrak g(x,y,\gamma^\mu)}
 \end{split}
 \end{equation}
with 
\begin{equation}
\begin{split}
\mathfrak g(x,y,\gamma^\mu) =& [(\gamma^1)^2 + (\gamma^3)^2]^2  \\ &+  (\gamma^2)^2 [(\gamma^3)^2 + (\gamma^1)^2]  +   (\gamma^0 \gamma^1 - \gamma^1 x)^2  \\ & +(\gamma^0 \gamma^3 - \gamma^3 x )^2 \\&- (  (\gamma^1)^2 +(\gamma^3)^2 +( \gamma^0 - x)^2) y^2 \\
  =& \text{Const}_1 - \text{Const}_2 y^2.
 \end{split}
 \end{equation}
Due to the positivity of these constants, the spectrum is gapless, and therefore the two EPs represent a surface state protected against being gapped out.

Now, if instead $\gamma^1 = 0$ and $\gamma^3 \neq 0$, we obtain
\begin{equation}
    \begin{split}
        k_y &= \frac{y}{\gamma^3} (x-\gamma^0) ,\\
        k_x^2 &= \left[ 1- \left( \frac{y}{\gamma^3} \right)^2 \right] (x - \gamma^0)^2 - y^2 + \vec \gamma \cdot \vec \gamma,
    \end{split}
\end{equation}
which can be solved for $(k_x, k_y)$ for
\begin{equation}
    \abs{y} \leq \frac{\sqrt{\left(x-\gamma^0\right)^2 + \vec \gamma \cdot \vec \gamma}}{\sqrt{1 + \left(x-\gamma^0\right)^2 / \left( \gamma^3 \right)^2}}.
\end{equation}
Therefore, the spectrum remains again gapless.

Lastly, we have to treat the case $\gamma^1 = \gamma^3 = 0$. The spectrum is then given by the equations
\begin{equation}
    \begin{split}
        y (\gamma^0 -x) &= 0,\\
        \left(\gamma^0 -x \right)^2 + \vec \gamma \cdot \vec \gamma - y^2 &= k_x^2 + k_y^2.
    \end{split}
\end{equation}
This means that the spectrum consists of the complete real axis and complex values at $x = \gamma^0$ with imaginary value bounded by $y_\text{max} = \abs{\vec \gamma}$. Such a spectrum is also gapless.
 
Thus, we have shown that a kinetic Hamiltonian of the form $k_x \sigma_x + k_y \sigma_z$ cannot be gapped out by addition of constant symmetric terms.

\section{Regularization of surface spectra for the ETI}

The exceptional topological insulator (ETI) emerges from a Hermitian three-dimensional topological insulator close to criticality when quasiparticles acquire a finite lifetime. This can be described on a cubic lattice of $s$ and $p$ orbitals by the Hamiltonian~\cite{denner2020exceptional}

\begin{equation}
\begin{split}
    H(\bs{k})&=\left(\sum_{j=x,y,z} \cos k_j-M\right) \tau_z\sigma_0\\
    &\quad+\lambda  \sum_{j=x,y,z}\sin k_j \, \tau_x\sigma_j+\mathrm{i}\delta\,\tau_x\sigma_0,
    \label{eq:ETI}
\end{split}
\end{equation}
where the Pauli matrices $\sigma_{\mu}$ and $\tau_{\mu}$ act on the spin and orbital degrees of freedom, respectively, with $\mu = 0, x, y, z$ and $\tau_0$ as well as $\sigma_0$ the $2\times2$ identity matrix. The parameter $M$ controls the band inversion between $s$ and $p$ orbitals and $\lambda$ the strength of spin-orbit coupling. A finite $\delta$ introduces the desired non-Hermiticity. For open boundary conditions, the model in Eq.~\eqref{eq:ETI} shows an infinitely steep set of eigenvalue branches, called infernal point. To highlight the generic surface state structure, we therefore add a small Zeeman term $\tau_0 \bs{B} \cdot \bs{\sigma}$ as a regularization, giving rise to a single surface band covering the central point gap.

\bibliography{Ref-Lib}